# Energy network: towards an interconnected energy infrastructure for the future

Haoyong Chen[1]*, Hailin Ge[1], Junzhong Wen[1], Ming Qiu[1], Hon-wing Ngan[2]

**Affiliations:**

[1]Department of Electrical Engineering, South China University of Technology, 510641, Guangzhou, China.

[2]Asia-Pacific Research Institute of Smart Grid and Renewable Energy, Hong Kong.

*Corresponding author. E-mail: *eehychen@scut.edu.cn* (H. Chen)

**Abstract:** The fundamental theory of energy networks in different energy forms is established following an in-depth analysis of the nature of energy for comprehensive energy utilization. The definition of an energy network is given. Combining the generalized balance equation of energy in space and the Pfaffian equation, the generalized transfer equations of energy in lines (pipes) are proposed. And the energy variation laws in the transfer processes are investigated. To establish the equations of energy networks, the Kirchhoff's Law in electric networks is extended to energy networks, which is called the Generalized Kirchhoff's Law. According to the linear phenomenological law, the generalized equivalent energy transfer equations with lumped parameters are derived in terms of the characteristic equations of energy transfer in lines(pipes). The equations are finally unified into a complete energy network equation system and its solvability is further discussed. Experiments are carried out on a combined cooling, heating and power(CCHP) system in engineering, the energy network theory proposed in this paper is used to model and analyze this system. By comparing the theoretical results obtained by our modeling approach and the data measured in experiments, the energy equations are validated.

## 0  Introductions

The ever increasing concern on energy crisis and pollution of environment has pushed forward the study on smart grid, renewable energy and techniques for enhancing energy efficiency such as the combined cooling, heating and power(CCHP) system [1], and the comprehensive utilization of various forms of energy has also become an important trend. Because the production, transmission and consumption of energy are often accomplished in the form of network, it forms the basis for conceptualizing the formation of the energy network such as one formed by the Britain and the Ireland founded Energy Network Association as the coordinating agency of electricity and gas distribution network [2]. The project "Vision of Future Energy Network" developed by Swiss Federal Institute of Technology has proposed the concept of energy hub and energy interconnector [3,4]. An energy hub is an integrated system of units in which multiple energy carriers can convert, adjust and store, while an energy interconnector is an underground device that enables the integrated transportation of electric energy, chemical energy and thermal energy. The proposal of these conceptions provides a new thought for modeling and analyzing the integrated energy network[5,6].

In the course of the study of energy network, there are many indispensable studies. For instance, Ref[7] deeply analyzes the coupling relationship between natural gas and electric power system in the United States, and puts forward the risk assessment method of power system considering operation constraint of natural gas pipeline. Ref [8] tries to investigate an optimal way to integrate the energy of both systems in urban areas so as to reduce wind curtailment, operation cost, and energy losses. Ref [9] introduces the concept of energy router, which adjusts dynamically the energy distribution in the network. It has put forward the architecture of energy network from a different perspective.

However, the most studies nowadays have not explored the unified physical mechanism of different forms of energy in an energy network in theory. With the development of distributed energy system and Energy Internet [10-12], the establishment of unified energy transfer and conversion laws and fundamental theory of integrated energy network are becoming more and more urgent. In the paper, a generalized approach is envisaged to model the generic nature of various kinds of energy transfer. By consolidating the basic theories of thermodynamics, heat transfer, electric network and fluid network, the generic attributes of energy are investigated. The generalized equations of energy transfer in lines (pipes) are derived based on the energy and its linear phenomenological law, and the Kirchhoff's laws are generalized for developing the basic equations of the energy network. To verify the above mentioned theories and approaches of energy network, experiments are carried out on a real engineering case of a typical energy network involving a CCHP system and the measured data are analyzed. This paper establishes the basic theory of energy network based on the deep discussion of energy essence, which lays an important foundation for modeling, analyzing and optimizing the integrated energy systems.

# 1 Fundamentals of an energy network

## 1.1 Basic structure of energy network

In engineering practice, energy is often provided to users through networks, such as electric power grids, heat supply networks and gas networks. Because a common feature of these networks is to supply energy, they are uniformly called energy network in this paper. Energy network consists of subnets with different forms of energy, and all these subnets are connected with each other through energy conversion devices. The relationship between the energy network and its subnets is analogous to the relationship between Internet and local area networks (LAN). The schematic diagrams of Internet and an energy network are shown in Fig. 1 and Fig.2 respectively. Energy can be transferred between the subnets of energy network, just as that information can be exchanged between the local area networks within the Internet. In Internet, different network protocols are often adopted in different LANs, and gateways are needed for protocol conversion and information exchange among different subnets of the Internet. This is also the case in energy network. The subnets of an energy network usually use different forms of energy, and conversion among these different energy forms is necessary, which is accomplished by energy conversion devices such as generators or electric heaters.

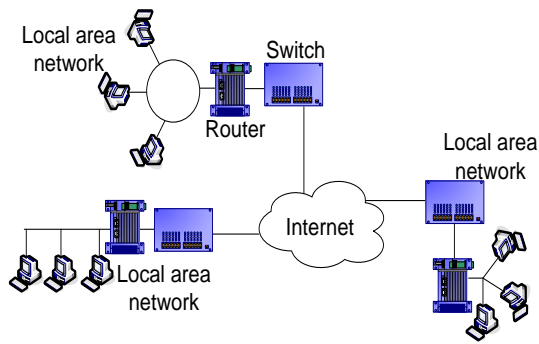
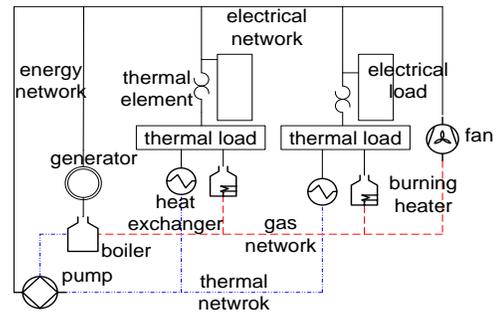

        **Fig.1 Diagram of Internet**        **Fig.2 Diagram of energy networks**

Although different types of energy carriers can't be mutually transformed in terms of substances, energy can be transferred from electricity to other substances or vice versa. Energy provides a link among different types of physical processes [13,14]. This is the fundamental that the energy network concept and theories rely on. A schematic diagram of chains of physical process is given in Ref[14] and redrawn in Fig. 3 to explain the underlying physical principles. The turbine converts hydro-power energy to mechanical energy; the generator converts mechanical energy to electric energy; the electric heater converts electric energy to thermal energy. The explanation of symbol meanings is left behind.

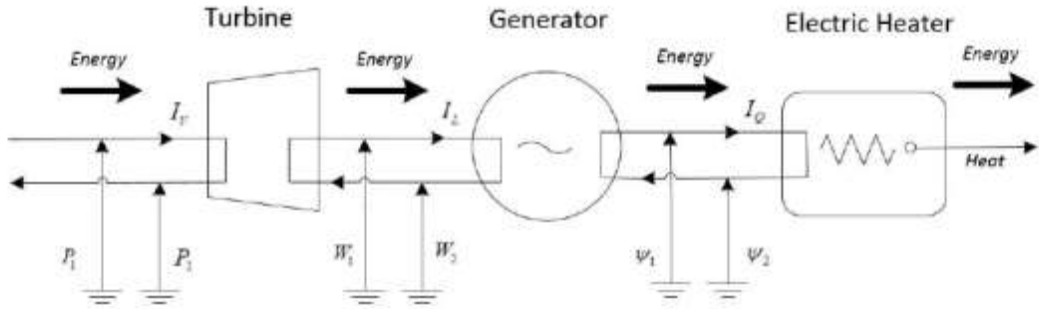

**Fig.3 The schematic diagram of chains of physical process**

## 1.2 Theoretical basis for the analysis of energy network

Energy network theory comes from the analysis for the process of energy transfer, providing a new method to study intergrated energy system. It generalizes various forms of matter into a unified form of intensive property and extensive property. Entropy, volume and quantity of charge can be treated as extensive properties while temperature, pressure and electric potential can be perceived as intensive properties. Based on the above definitions, the transfer equations for each sub-network are deduced. Before that, the theoretical basis for the mathematical derivation of the energy network transfer equation should be introduced:

### 1.2.1 *Kirchhoff's intensive property law (KIL) and Kirchhoff's extensive property law (KEL)*

Based on the current study of the models of the electrical network, thermal networks and gas networks, we can observe that each network above is in line with Kirchhoff's first law and the Kirchhoff's second law[15,16,17]. Therefore, we might as well assume that Kirchhoff's laws can be extended to the entire intergrated energy network. Then, the generalized Kirchhoff theory is put forward. In correspondence with Kirchhoff's voltage law (KVL) and Kirchhoff's current law (KCL) in electrical network, we propose Kirchhoff's intensive property law (KIL) and Kirchhoff's extensive property law (KEL) in energy network:

(1) Kirchhoff's intensive property law (KIL): in any circuit of an energy network, the sum of intensive property difference in each transfer line (pipe) is 0, and this law can be described by $\sum \Delta X = 0$.

(2) Kirchhoff's extensive property law (KEL): in an energy network, the extensive property that flows into any node is equal to the extensive property that flows out of the node.

With the generalized Kirchhoff's Law, the equations of energy network can be established just like the equations of electric network. Based on the generalized balance equations of extensive property and energy in space, the generalized equations of energy transfer in lines (pipes) are further derived. Hence, a key step of the paper is to develop an equivalent model of energy transfer in lines (pipes) through the analysis of the energy transfer mechanism, derive the exergy transfer rule and establish the energy network equations.

### 1.2.2 *Pfaffian expression of the external work*

When we describe the energy transfer equation in the transmission lines (pipes), the energy postulate of the whole energy system should be firstly introduced. According to the theoretical basis of axiomatic thermodynamics proposed by C. Caratheodory[18], all thermodynamic equations, including the first law of thermodynamics, can be expressed as the form of a Pfaffian equation. Thus, the external work $W$ is calculated as follow:

$$dW = \sum_{i=1}^{n} p_i dx_i \tag{1}$$

Where $x_i$ is the state variables of a system and $p_i$ is the only function of $x_i$. The same expression can be found in Ref [19]. And we extend the equation (1) to the field of integrated energy system:

$$dE = \sum_{i=1}^{n} x_i dX_i \tag{2}$$

While $E$ donates the energy loss caused by intensive property $X$ promoting the flow of extensive property $x$. In above equation, each pair of parameters is independent. Therefore, all the parameters are limited by:

$$\begin{cases} X_i = \dfrac{\partial E}{\partial x_i} \\ \dfrac{\partial X_i}{\partial x_j} = \dfrac{\partial X_j}{\partial x_i} = 0 \end{cases} \tag{3}$$

The equations above are the basis for our theoretical derivation.

## 2 The generalized equations of energy transfer

### 2.1 The generalized equations of energy transfer

Essentially, energy transfer is realized through the transfer of related extensive property[13,14,18]. Hence, it is necessary to study the transfer laws of extensive property in the energy transfer process.

According to the mass conservation equation in Ref[20], the generalized balance equation of extensive property in space is derived.

$$\rho \frac{dx_i}{dt} = -\nabla \cdot \boldsymbol{J}_{xi} + g_{xi} \tag{4}$$

Combining the Pfaffian expression (2) of energy and the equation (4), the energy conservation equation of extensive property and intensive property is shown as follow:

$$\rho \frac{dX_i \cdot x_i}{dt} = -\nabla \cdot (X_i \boldsymbol{J}_{xi}) + X_i g_{xi}$$

$$\Rightarrow \rho \frac{X_i dx_i}{dt} = -\nabla \cdot (X_i \boldsymbol{J}_{xi}) + \boldsymbol{J}_{xi} \cdot \nabla X_i + X_i g_{xi}$$

(5)

Where $\rho$ is the density of medium; $x_i$ is the extensive property in each unit mass of medium; $X_i$ is the intensive property; $g_{xi}$ is the source strength of $x_i$ in each unit volume of medium; $\boldsymbol{J}_{xi}$ is the flow density vector of $x_i$, which means the amount of $x_i$ that flows through each unit area per unit time. And the equation (5) is the same as the energy conservation equation in Ref[20].

In engineering practice, lines (pipes) for energy transfer often take a cylindrical shape, such as electrical lines, heat-supply pipes and gas pipelines. The section shape of lines (pipes) only affects the flow resistance value of the lines (pipes), and this means that the lines (pipes) with a fixed section shape have a certain flow resistance value per unit length. In this paper, the building block of an energy network is based on the cylindrical transfer lines (pipes) as shown in Fig.4. The intensive property gradient only exists in the axial direction, so energy is transferred in one-dimension.

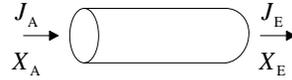

**Fig.4 Cylindrical transfer lines (pipes)**

By calculating volume integral for both sides of equation(4), we get

$$\int_v \left( \rho \frac{dx}{dt} \right) dv = \int_v (-\nabla \cdot \boldsymbol{J}_x) dv + \int_v g_x dv$$

$$\Rightarrow \frac{dx^*}{dt} = -\oint_A \boldsymbol{J}_x \cdot d\boldsymbol{A} + \int_0^L x_g dL$$

(6)

where $x^*$ is the total extensive property in the cylindrical transfer line (pipe); $x_g$ is a function of the extensive property generated in transfer process with regard to line (pipe) length $L$; the subscript $i$ in equation(4) is omitted.

As the extensive property $x$ just flows inside the cylindrical transfer line (pipe), and there is no $x$ flowing through the surface of the cylindrical transfer line (pipe) while the direction of area $A$ is its normal direction, equation (6) can then be further simplified as:

$$\frac{dx^*}{dt} = \sum JA + x_g = H_A - H_E + \int_0^L x_g dL$$

(7)

In equation(7), $H$ is the amount of the extensive property of $x$ that flows through the sectional area $A$ per unit time. Equation(7) is the balance equation of extensive property in transfer process along the cylindrical transfer line (pipe). Hence, equation(7) shows that, in unit time, the variation of $x$ in a transfer line (pipe) is equal to the sum of $x$ that flows into the line (pipe) through the sectional area $A$ and $x$ that is generated flowing along the line (pipe).

In the same way, calculating volume integral for both sides of equation(5) yields:

$$\int_v \rho \frac{Xdx}{dt}dv = \int_v -\nabla \bullet (X\vec{J}_x)dv + \int_v \vec{J}_x \nabla X dv + \int_v Xg_x dv$$
$$\Rightarrow \frac{dE}{dt} = -\oint_A (X\vec{J}_x)d\vec{A} + \int_A^E \vec{J}_x \frac{dX}{dL}AdL + \int_A^E Xdx_g \tag{8}$$

$E$ is the total energy in the cylindrical transfer line (pipe).

As the intensive property $X$ just varies along the axial direction and keeps unchanged along the radial direction while there is no energy flowing through the surface of the cylindrical transfer line (pipe), equation(8) can be further simplified as:

$$\frac{dE}{dt} = \sum XH + \int_A^E HdX + \int_A^E Xdx_g = X_A H_A - X_E H_E + \int_A^E HdX + \int_A^E Xdx_g \tag{9}$$

In equation(9), the expression on the left is the variation of energy in the cylindrical transfer line (pipe) per unit time. The first and second term on the right is the energy that flows into the boundary of the cylindrical transfer line (pipe); the third one is the consumption of exergy in the transfer process; the fourth one is the energy that emerges with the generation of new extensive property.

Equations(7) and (9) are the general time-variant equations of transfer process in the cylindrical line (pipe). For the time-invariant equations, terms on the left of equation (7) and equation (9) are both 0, namely the energy and extensive property in the transfer line (pipe) remain unchanged. In this situation all extensive property generated will flow out, thus

$$dx_g = dH \tag{10}$$

The time-invariant balance equations of energy and extensive property in the transfer line (pipe) are given in equation(11) and equation(12):

$$H_A - H_E + \int_0^L x_g dL = 0 \tag{11}$$

$$X_A H_A - X_E H_E + \int_A^E HdX + \int_A^E XdH = 0 \tag{12}$$

**2.2 Analysis of the energy transfer process**

Generally, energy in an integrated energy system can converts with each other. To describe the energy state of a system, multiple pairs of intensive and extensive properties are often needed. In this paper, to analyze the transfer law of one energy form, the other energy forms and their corresponding intensive and extensive properties are assumed to be unchanged in the energy transfer process.

The variation of extensive property and intensive property in an energy transfer process of the line (pipe) shown in Fig. 4 can be illustrated with Fig. 5. According to equation(12), the absolute value of the exergy loss in this energy transfer process is area $S_{AEGD}$, and $\int_{A}^{E} HdX = -S_{AEGD}$; the energy increase due to the increase of extensive property in the transfer process is area $S_{AENM}$, and $\int_{A}^{E} XdH = S_{AENM}$. Thus the total energy loss in the transfer process is $S_{AENM} - S_{AEGD}$. We have

$$\int_{A}^{E} HdX + \int_{A}^{E} XdH = S_{AENM} - S_{AEGD} = S_{ENMF} + S_{AEF} - (S_{AFGD} + S_{AEF}) \\ = S_{ENOG} - S_{AMOD} = X_E H_E - X_A H_A \quad (13)$$

which is just the relationship of equation(12).

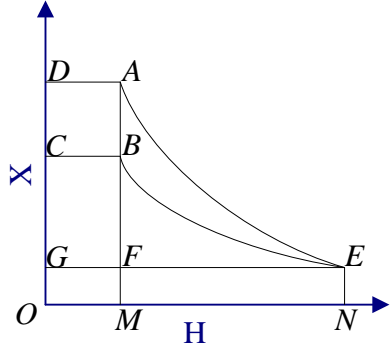

**Fig.5 Change curve of extensive property - intensive property**

The energy difference between state A and state E on curve AE represents the energy loss of the system. According to Pfaffian expression, energy is a function of the state parameters $X$ and $H$, which is irrelevant to the energy transfer process. Hence the energy transfer in the process AE is equal to that of the process ABE. The process AB denotes a transfer process in which the extensive property is kept unchanged, such as electric energy transfer; the process BE denotes a transfer process in which the energy is kept unchanged, such as constant temperature heat conduction.

Taking the gas expansion process for instance, if the process is thermostatic, the expansion process is represented by curve BE in Fig.5. In this process, the total pressure energy of gas remains constant as the heat absorbed is entirely converted to expansion work. If the process is not thermostatic, there is energy exchange between pressure energy and thermal energy in the expansion process, and a part of the heat absorbed will be converted to pressure energy.

## 3 The equivalent transfer equation of lumped element model

The original Kirchhoff's circuit laws are two equalities that deal with the current and potential difference (voltage) in the lumped element model of electrical circuits. The KIL and KEL of energy network also depend on the lumped element models being applicable. The characteristic equation of transfer line (pipe) is the equation describing the relationship between intensive property and extensive property in line (pipe), and the generalized characteristic equation of transfer lines (pipes) is derived here.

### 3.1 *Linear phenomenological law*

According to the Fourier law, ref[20] points out the linear phenomenological law that all kinds of flows are promoted by the force in the system, and deduces their relationship:

$$J_i = \sum_{i=1}^{n} L_{ij} F_j \tag{14}$$

Where $J_i$ donates the flow, $F_j$ donates the force and $L_{ij}$ is the phenomenological coefficient. As our single-energy flow subnet, there is only one kind of flow and force, equation (14) can be extended and simplied into:

$$\boldsymbol{J}_{xi} = K_i \boldsymbol{F}_i = -K_i \nabla X_i \tag{15}$$

Thus, the expression of extensive property flow $H$ is equation (16).

$$H = J_x A = -KA \nabla X \tag{16}$$

Where the subscript in equation(15) is omitted.

In (16), $A$ is the sectional area through which the extensive property flows.

### 3.2 *The equivalent transfer equation*

By integrating equation(16), we can obtain an equation expressing the relationship between the extensive and intensive property. Since $K$ may or may not be a constant value, the two different situations should be discussed separately.

(a) If $K$ is a constant value, such as the transfer process of electric energy and the transfer process of pressure energy with steady incompressible laminar flow, etc. By integrating equation(16), we get

$$H = -KA \frac{dX}{dL} \Rightarrow \int_A^E \frac{H}{KA} dL = \int_A^E -dX \Rightarrow \frac{1}{KA} \int_A^E H dL = X_A - X_E$$
$$\Rightarrow H^* R = X_A - X_E \quad R = \frac{L}{KA} \tag{17}$$

where $H^*$ is the Lagrange Mean Value of integral $\int_A^E H dL$, which is called the equivalent extensive property flow in the transfer process, and its value is decided by the intensive property difference between the two ends of a line (pipe) and the constant value $R$. $R$ characterizes the resistance in the energy transfer process, herein generally referred to as energy resistance. $R$ is the electrical resistance in the transfer process of electric energy, and it is the flow resistance in the transfer process of pressure energy.

(b) If $K$ is not a constant value, while the energy transfer coefficient $\lambda$ ($\lambda = KA$) is constant, such as the heat conduction process, we have

$$H = -\frac{KXA}{X}\frac{dX}{dL} \Rightarrow \int_A^E \frac{H}{\lambda A}dL = \int_A^E -\frac{1}{X}dX \Rightarrow \frac{1}{\lambda A}\int_A^E HdL = \ln X_A - \ln X_E$$
$$\Rightarrow H^*R = \ln X_A - \ln X_E \quad R = \frac{L}{\lambda A}$$
(18)

where $H^*$ is still the equivalent extensive property flow. However, the relationship between $H$ and $X$ is nonlinear in this case. In fact, $H$ is linear with respect to $\ln X$. As for the transfer process that $K$ and $\lambda$ are both not constant, it can be equivalent to transfer processes with constant $K$ and $\lambda$ introduced in Section 2.2.

According to the above discussion, the equivalent extensive property flow $H^*$ can be used to describe the transfer process in lines (pipes). In this way, the integral equation can be turned into algebraic equation, which is easier to be solved. The variation of energy in the equivalent transfer process is shown in Fig.6.

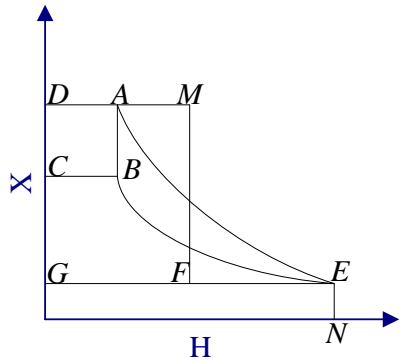

**Fig.6 The equivalent transfer process diagram**

Curve $AE$ in Fig. 6 is the original transfer process. Because energy is a function of state, the variation of one energy form in a system is only related to its corresponding state parameters rather than the specific process of variation. The difference of energy between two states is just the value of energy variation in the system. Thus, the variation value of energy in process $AE$ is equal to that in process $ABE$, and also equal to that in process $AMFE$. Supposing $H_{MF} = H^*$ in Fig. 6, which means the equivalent extensive property flow of equivalent transfer process $AMFE$ is $H^*$, this process can be described by using lumped element model.

The equivalent physical model of the process $AMFE$ is shown in Fig.7.

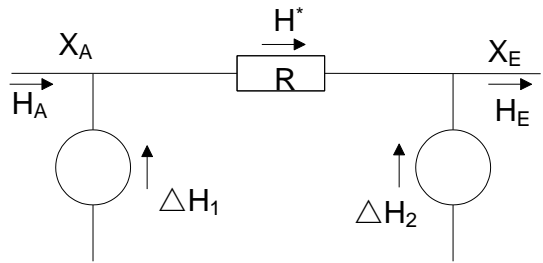

**Fig.7 The physical model of equivalent transfer**

With respect to the transfer process of electric energy and pressure energy in steady incompressible laminar flow, the current and the fluid flow keeps constant in the transfer process, hence $\Delta H_1 = \Delta H_2 = 0$, and a linear relationship between $H$ and $X$ can be obtained:

$$H^* = H_1 = H_2 = \frac{X_1 - X_2}{R} \tag{19}$$

In the electric energy transfer process, equation(19) is just the Ohm's Law, where $R$ is the resistance, $H^*$ is the electric current.

In the transfer process of steady incompressible laminar flow, supposing the transfer coefficient of volume $K_v$ to be constant and defining the flow resistance as $R_v = \frac{L}{K_v A}$, then equation $H_v = \frac{X_1 - X_2}{R_v}$ can be obtained. The flow resistance of the laminar flow in the cylindrical transfer line (pipe) is [13,14]:

$$R_v = \frac{128 \mu L}{\pi D^4} \tag{20}$$

In equation(20), $\mu$ is the dynamic viscosity of fluid, which is related to temperature and fluid type. Compared with the definition of flow resistance in equation(20), the transfer coefficient of volume is $K_v = \frac{D^2}{32\mu}$. $K_v$ is constant for a certain fluid flowing in a certain transfer line (pipe), and this conclusion accords with the previous assumption.

Considering the constant temperature heat conduction in thermal energy transfer, in which case $K$ is not a constant value, but the heat transfer coefficient $\lambda$ ($\lambda = KX$) is constant, the following equation can be obtained according to equation(18):

$$H^* R = \ln X_A - \ln X_E \quad R = \frac{L}{\lambda A} \tag{21}$$

In equation(21), $H^*$ is the equivalent to the entropy flow, while $X$ denotes temperature. In constant temperature heat conduction process, heat flow is constant and the equation can be further deduced as follows:

$$P = J_{en} A = X J_x A = -KXA \nabla X = -\lambda A \frac{dX}{dL}$$
$$\Rightarrow \frac{P}{\lambda A} dL = -dX \quad \Rightarrow P = \lambda A \frac{X_1 - X_2}{L} \tag{22}$$

Equation(22) is just the equation of one-dimensional constant temperature heat conduction in heat transfer theory.

There are various forms of energy used in engineering practice, but the energies transferred in network are mainly electric energy and thermal energy. Other forms of energy such as nuclear energy, wind energy, solar energy and hydropower energy are converted into electric energy or thermal energy before it is transferred. Thus the equation(19) is the key to model an energy network.

# 4 Analysis of a typical energy network

## 4.1 Formulation of the equation system of a typical energy network

The system formed by a combined cooling, heating and power (CCHP) generator in engineering practice is a typical energy network. In this section, a real CCHP system is modeled and analyzed, and experiments are carried out to validate the theoretical results.

The major equipment of this CCHP system includes: gas turbine, absorption-type refrigerating and electric refrigeration units, pump and so on. The electricity produced by the gas turbine is supplied to a multiple-use building, a laboratory building and a new building, and the shortage of electricity is complemented by an outside power grid. On the other hand, the cooling energy produced by absorption-type refrigerating unit is transferred to the multiple-use building and the new building, the lack of cooling energy is complemented by electric refrigeration. The system configuration diagram is shown in Fig.8.

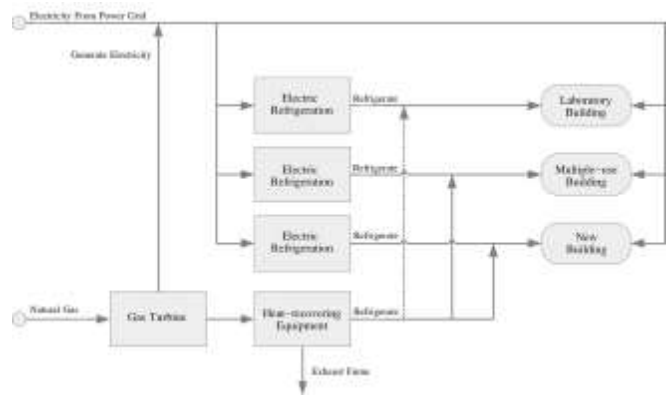

**Fig.8 CCHP system configuration diagram**

To facilitate modeling and analysis of the system, this paper commences the study on the system at 12pm on September 2, 2010. At the moment, since the absorption-type refrigerating unit provides cold energy to the new building only, the multiple-use building and the laboratory building are excluded in the model. As gas turbine is connected to the main power grid, and then electricity supply of the system can be regarded as adequate; supply from the gas turbine and power grid can be regarded as the same electric energy source. According to the above configuration, the energy network diagram of the CCHP system is depicted as Fig.9.

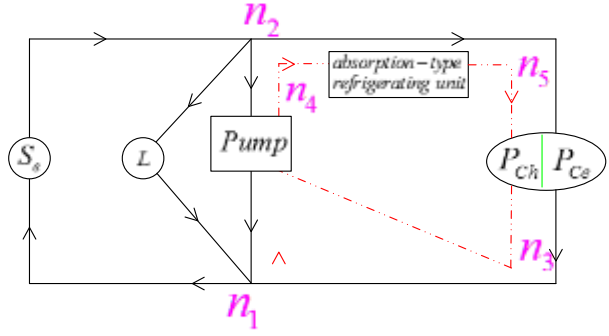

**Fig.9 Energy network diagram of CCHP system**

In Fig.9, the solid lines denote the transfer route of electric energy within the electric network; dot dash lines denote the transfer route of cooling energy, which is carried by the water flow in the pipes of water network driven by pressure energy. Furthermore, electric network is linked with water network through pump and cooling load ($P_C = P_{Ch} + P_{Ce}$, where $P_{Ch}$ is the cooling power supplied by the absorption-type refrigerating unit; $P_{Ce}$ is the cooling power supplied by the electric refrigeration units; $P_C$ is the total cooling power); users get the cooling energy through the heat exchanger. Although the cooling energy exists in the energy network, it has not yet formed a transfer network of its own (the transfer of cooling energy is implemented through water flow in the water network). $S_e$ is electric energy source; $L$ is other load of electric network; absorption-type refrigerating unit is the cooling energy source of the water network.

According to the basic network topology of the CCHP system shown in Fig.9, the nodes, braches and network topologies can be simplified as Fig.10.

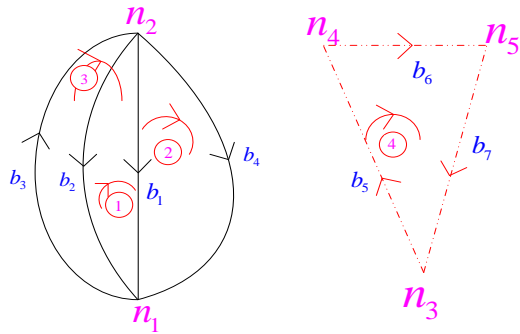

**Fig.10 Simplified network diagram of CCHP system**

Choosing $n_1$, $n_3$ as reference node, the correlation matrix of this network can be written as

$$A = \begin{bmatrix} 1 & 1 & -1 & 1 & 0 & 0 & 0 \\ 0 & 0 & 0 & 0 & -1 & 1 & 0 \\ 0 & 0 & 0 & 0 & 0 & -1 & 1 \end{bmatrix}$$

The extensive property flow vector is:

$$\boldsymbol{H} = [H_1 \ H_2 \ H_3 \ H_4 \ H_5 \ H_6 \ H_7]^T$$

The basic circuit matrix is:

$$B_f = \begin{bmatrix} 1 & -1 & 0 & 0 & 0 & 0 & 0 \\ -1 & 0 & 0 & 1 & 0 & 0 & 0 \\ 1 & 0 & 1 & 0 & 0 & 0 & 0 \\ 0 & 0 & 0 & 0 & 1 & 1 & 1 \end{bmatrix}$$

The vector of intensive property difference between the two ends of lines (pipes) is:

$$\Delta X = [\Delta X_1 \ \Delta X_2 \ \Delta X_3 \ \Delta X_4 \ \Delta X_5 \ \Delta X_6 \ \Delta X_7]^T$$

According to the Generalized Kirchhoff's Law, the system of equations of this energy network can be written as:

$$\begin{cases} AH = 0 \\ B_f \Delta X = 0 \end{cases}$$

In this equation system, there are 7 equations and 14 unknown variables. Thus another 7 equations are required to solve this equation group. In Fig.10, all branches, except the electric energy source branch $b_3$, the pump branch $b_5$, the branch $b_1$ showing electric energy transfer to pump and the electric cooler branch $b_4$, can be described by the characteristic equations of transfer lines(pipes): $H_i = \dfrac{\Delta X_i}{R_i}$ ($i = 2, 6, 7$), $R_i = \dfrac{L_i}{K_i A_i}$. Thus, 3 equations can be obtained. According to the energy conservation law, an equation describing energy transfer between pump branch $b_5$ and branch $b_1$: $\Delta X_5 H_5 + \eta \Delta X_1 H_1 = 0$ ($\eta$ is energy conversion efficiency ratio of pump) is obtained. The equation of electric refrigeration branch $b_4$ can be written as $P_{Ce} = COP \times \Delta X_4 H_4$. already have 12 equations.

Usually, the voltage of electric energy source branch ($\Delta X_3$) is known. If $P_{Ch}$ is known, $H_7$ can be solved through equation $P_{Ch} = \rho c \Delta T H_7$. Thus, the number of unknown variables in this energy network equation system becomes 12. With the above 12 equations, the equation system can be solved.

### 4.2 Numerical analysis of the energy network

Based on the theoretical analysis of the typical energy network in Section 4.1, the practical data of the CCHP system is used to analyze the energy network. When the total cooling load, voltage of the electric energy branch and parameters of lines (pipes) are known, the operating parameters of the pump and electric energy source can be solved.

For the typical energy network in Fig.9, the known parameters include: total cooling load $P_C = 875.18 KW$, cooling power of the absorption-type refrigerating unit $P_{Ch} = 334.64 KW$, and the cooling power of electric refrigeration unit $P_{Ce} = 540.54 KW$; the voltage of electric energy branch $\Delta X_3 = -230.19V$,

$R_2 = R_L = 2.15\Omega$, $R_6 = 2.4\times 10^6\, Pa\cdot s/m^3$, $R_7 = 3.6\times 10^6\, Pa\cdot s/m^3$; $R_4$ is a variable resistance whose value affects the cooling power of electric refrigeration; the energy conversion efficiency of pump $\eta = 0.52$, the energy efficiency of electric refrigeration $COP = 4$. The solving steps include:

(1) Establishing the equation system of energy network
$$\begin{cases} AH = 0 \\ B_f \Delta X = 0 \end{cases}$$

(2) Formulating the characteristic equations of lines (pipes):
$$\Delta X_i = H_i R_i \quad (i = 2, 6, 7)$$

(3) Formulating the energy conservation equation
$$\Delta X_5 H_5 + \eta \Delta X_1 H_1 = 0$$

(4) Formulating the equation of electric refrigeration branch
$$P_{Ce} = COP \times \Delta X_4 H_4$$

(5) Solving the extensive property flow $H_7$

As the cooling power of the absorption-type refrigerating unit $P_{Ch} = 334.64 KW$ is already known, according to equation $P_{Ch} = \rho c \Delta T H_7$, the fluid flow $H_7 = 3.331\times 10^{-2}\, m^3/s$ can be obtained.

(6) Solving the equation system

As the units of voltage and pressure used in physics and engineering are different, the values of the two parameters can not be compared directly. Also in order to simplify the calculation, this paper chooses the per-unit system used in electric power system analysis to solve the equation system, in which the system properties are expressed as fractions of a defined base unit quantity[21]. There are two kind of energy subnet in this system including electric network and thermal network. Correspondingly, the base intensive properties in this system are voltage ($\Delta X_1, \Delta X_2, \Delta X_3, \Delta X_4$) and pressure ($\Delta X_5, \Delta X_6, \Delta X_7$) while the extensive properties are current ($H_1, H_2, H_3, H_4$) and fluid flow ($H_5, H_6, H_7$). The choice of base intensive property and extensive property should satisfy the energy conservation law.

The following base properties are chosen in this paper: fluid flow base value $H_{PB} = 10^{-3}\, m^3/s$, pressure base value $\Delta X_{PB} = 10^3\, Pa$, voltage base value $\Delta X_{qB} = 230V$. Thus according to the energy conservation law, the current base value should be $H_{qB} = H_{PB} \times \Delta X_{PB} / \Delta X_{qB} = 1/230\, A$. We can future calculate fluid flow resistance base value $R_{PB} = 10^6\, Pa\cdot s/m^3$ and electrical resistance base value $R_{qB} = 5.29\times 10^4\, \Omega$. The per-unit values of the following practical data are calculated as:

Total cooling load:
$$P_{Cs} = \frac{8.7518\times 10^5\, w}{10^3 \times 10^{-3}\, w} = 8.7518\times 10^5$$

Cooling power supplied by the water network:

$$P_{Chs} = \frac{3.3464 \times 10^5 w}{10^3 \times 10^{-3} w} = 3.3464 \times 10^5$$

Cooling power supplied by the electric network:

$$P_{Ces} = \frac{5.4054 \times 10^5 w}{10^3 \times 10^{-3} w} = 5.4054 \times 10^5$$

Voltage:

$$\Delta X_{3s} = \frac{-230.19V}{230V} = -1.001$$

Fluid flow:

$$H_{7s} = \frac{3.331 \times 10^{-2} m^3/s}{10^{-3} m^3/s} = 33.31$$

Resistance:

$$R_{2s} = \frac{2.15 \Omega}{5.29 \times 10^4 \Omega} = 4.064 \times 10^{-5}$$

Fluid flow resistance:

$$R_{6s} = \frac{2.4 \times 10^6 Pa \cdot s/m^3}{10^6 Pa \cdot s/m^3} = 2.4$$

$$R_{7s} = \frac{3.6 \times 10^6 Pa \cdot s/m^3}{10^6 Pa \cdot s/m^3} = 3.6$$

Solving the equation system, we can obtain:

$$\Delta X_s = \begin{bmatrix} 1.001 & 1.001 & -1.001 & 1.001 & -199.86 & 79.944 & 119.916 \end{bmatrix}$$

$$H_s = \begin{bmatrix} 12789.8 & 24630.9 & 172420.7 & 135000 & 33.31 & 33.31 & 33.31 \end{bmatrix}$$

By multiplying $\Delta X_s$ and $H_s$ by the corresponding base values, we can obtain:

$$\Delta X = \begin{bmatrix} 230.23 & 230.23 & -230.23 & 230.23 & -199860 & 79944 & 119916 \end{bmatrix}$$

$$H = \begin{bmatrix} 55.608 & 107.091 & 749.656 & 586.957 & 0.03331 & 0.03331 & 0.03331 \end{bmatrix}$$

If the intensive properties of reference nodes ($n_1$, $n_3$) are known, the intensive properties of all nodes in energy network can be obtained. If the voltage of $n_1$ is its dead state value $0\ V$, the pressure of $n_3$ is its dead state value $10^5 Pa$, then the intensive properties of the nodes are:

$$X = \begin{bmatrix} 0 & 230.23 & 100000 & 299860 & 219916 \end{bmatrix}^T$$

### 4.3 Experimental verification of the energy network equations

Through the numerical analysis of the energy network, the calculated operating parameters of the CCHP system at a specific time are obtained. Then we can compare them with the data measured in the experiment. The vectors of intensive property difference and extensive property in the experiment on CCHP system are as follows:

$$\Delta X_a = \begin{bmatrix} 230.19 & 230.19 & -230.19 & 230.19 & -200000 & 80000 & 120000 \end{bmatrix}$$
$$H_a = \begin{bmatrix} 58.26 & 107.087 & 770.565 & 605.2 & 0.03322 & 0.03322 & 0.03322 \end{bmatrix}$$

The actual and calculated operating parameters are listed in Table1, with the errors showing the difference between the actual and calculated operating parameters. The results show that absolute values of the errors are all within 5%. Since the analysis of the energy network is based on a steady state model, while the practical operating process is dynamic, the error between the calculated and actual data is acceptable. It implies that the energy network theories and equations are correct.

Table.1 Comparison of operation parameters

| Parameters | Actual operating parameters | Calculated operating parameters | Error |
|---|---|---|---|
| $\Delta X_1$ | 230.19 | 230.23 | 0.02% |
| $\Delta X_2$ | 230.19 | 230.23 | 0.02% |
| $\Delta X_3$ | -230.19 | -230.23 | 0.02% |
| $\Delta X_4$ | 230.19 | 230.23 | 0.02% |
| $\Delta X_5$ | -200000 | -199860 | -0.07% |
| $\Delta X_6$ | 80000 | 79944 | -0.07% |
| $\Delta X_7$ | 120000 | 119916 | -0.07% |
| $H_1$ | 58.26 | 55.608 | -4.55% |
| $H_2$ | 107.087 | 107.091 | 0.004% |
| $H_3$ | 770.565 | 749.656 | -2.71% |
| $H_4$ | 605.2 | 586.957 | -3.01% |
| $H_5$ | 0.03322 | 0.03331 | 0.27% |
| $H_6$ | 0.03322 | 0.03331 | 0.27% |
| $H_7$ | 0.03322 | 0.03331 | 0.27% |

## 5 Conclusion

To cope with energy crisis and climate change, comprehensive utilization of energy in different forms has become an inevitable trend. In engineering practice, the production, transmission and utilization of energy are often accomplished by the means of energy networks such as electric power grid, heat supply network, gas network, etc. However, there is still no theory for general energy networks containing different forms of energies. The fundamental theory of energy networks is established following an in-depth analysis of the nature of energy and by consolidating the basic theories of thermodynamics, heat transfer, electric network and fluid network.

The generalized transfer equations of energy in lines (pipes) are proposed based on the generalized balance equation of energy in space, and the energy and exergy variation laws in the transfer processes are investigated. Because energy is a state quantity, the variation of energy in a system is only relevant to the intensive and extensive properties at the initial and final states instead of the specific transfer process. In case that the initial and final states of different transfer processes are identical, the variations of energy transfer in these processes are equivalent. Then the generalized equivalent energy transfer equations with lumped parameters are derived on basis of the characteristic equations of energy transfer in lines (pipes).As in the electric network theory, the equations are finally unified into a complete energy network equation system by using generalized Kirchhoff's laws and its solvability is further discussed.

Experiments are carried out on a CCHP system in engineering, the energy network theory proposed in this paper is used to model and analyze this system. By comparing the theoretical results obtained by our modeling approach and the data measured in experiments, the energy equations are validated.